\documentclass[prl,twocolumn,showpacs,superscriptaddress]{revtex4}

\usepackage{graphicx,amsmath,amssymb,psfrag,color}
\usepackage{dcolumn}
\usepackage{bm}

\newcommand{\be}{\begin{equation}} 
\newcommand{\ee}{\end{equation}}
\newcommand{\bea}{\begin{eqnarray}} 
\newcommand{\eea}{\end{eqnarray}}

\begin{document}

\title{Surface acoustic wave propagation in monolayer graphene}


\author{Peter Thalmeier}
\affiliation{Max-Planck-Institut f\"ur Chemische Physik fester Stoffe, 01187 Dresden, Germany}
\author{Bal\'azs D\'ora}
\email{dora@pks.mpg.de}
\affiliation{Max-Planck-Institut f\"ur Physik komplexer Systeme, N\"othnitzer Str. 38, 01187 Dresden, Germany}
\author{Klaus Ziegler}
\affiliation{Institut f\"ur Physik, Universit\"at Augsburg, D-86135 Augsburg, Germany}

\date{\today}

\begin{abstract}
Surface acoustic wave (SAW) propagation is a powerful method to investigate 2D electron systems. We show how SAW observables are 
influenced by coupling to the 2D massless Dirac electrons of graphene and argue that Landau oscillations can be observed as function 
of gate voltage for constant field. Contrary to other transport measurements, the zero-field SAW propagation gives the wave 
vector dependence of graphene conductivity for small wave numbers. We predict a crossover from Schr\"odinger to Dirac like behaviour 
as a function of gate voltage, with no attenuation in the latter for clean samples.
\end{abstract}



\maketitle

\begin{figure}
\vspace{0.2cm}
\includegraphics[width=75mm]{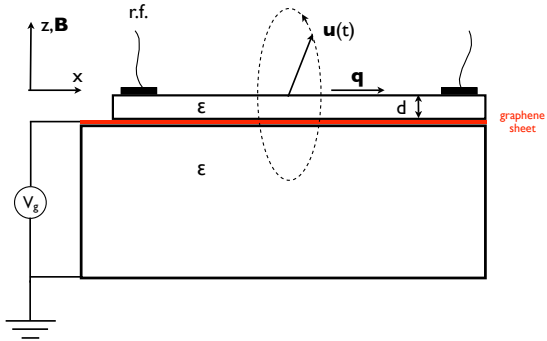}
\caption{Schematic setup for SAW propagation on monolayer graphene sandwiched between piezoelectric slabs with dielectric constant 
$\epsilon$ as adapted from Ref.~\onlinecite{Simon96}. SAW transducer and detector are attached to the top slab with thickness 
$qd\ll1$ ($q$ = SAW wave number ). The SAW displacement ${\bf u}(t)$ at fixed position traces an ellipse in the sagittal ($xz$) plane 
and the penetration depth in -z direction is $\sim 2\pi/q \gg d$.  Direction of field {\bf B} is normal to the surface. }
\label{fig:Fig1}
\end{figure}

Sound waves in solids may be used as diagnostic tool to investigate the low frequency long wavelength response of electrons. In metals 
they couple via the deformation potential mechanism, i.e. stress dependence of kinetic energy. In semiconductors the piezoelectric 
mechanism which generates internal electric fields that accelerate the charge carriers is most effective provided the compound has a 
piezoelectrically active structure. Close to smooth surfaces the acoustic modes may become localized in the direction of the surface 
normal and propagate as surface acoustic waves (SAW). The most prominent ones are Rayleigh waves \cite{Stoneley55} where the displacement
 vector ${\bf u}(t)$ lies in the 'sagittal plane' formed by the surface normal vector $\hat{{\bf n}}$ and the acoustic propagation vector
 $\bf q$ which is parallel to the surface (normal to $\hat{{\bf n}}\equiv \hat{{\bf z}}$, cf. Fig.~\ref{fig:Fig1}) . The Rayleigh 
SAW may
 be thought as consisting of longitudinal and transverse components whose amplitudes depend on the normal direction in a manner that the 
surface fulfills the mechanical stress free boundary condition. The Rayleigh wave is therefore an elliptically polarized wave in the 
sagittal plane which decays exponentially into the bulk with an acoustic penetration depth that is of the order of the bulk wavelength. 

Because of these properties the Rayleigh SAW exhibit interesting effects due to the coupling to charge carriers, particularly in the 
presence of a magnetic field. Two of those are well known and have been explored experimentally. i) when the field is within the surface 
plane and perpendicular to the sagittal ($xz$) plane then due to the breaking of time reversal invariance the velocity of propagation  
of SAW with wave vectors {\bf q} and {-\bf q} is different. This non-reciprocal SAW effect has been investigated for metals in detail 
in Ref.~\onlinecite{Heil84,Luethibook}. ii) In semiconductor heterostructures the charge carriers are confined in the lateral ($xy$) plane 
and a 2D electron gas (2DEG) forms. This may be nicely investigated by Rayleigh SAW which are localized on the surface sheet. For fields 
perpendicular to the surface they probe the longitudinal wave vector dependent conductivity  of the 2DEG. Since the latter exhibits the 
typical Landau quantum oscillations under suitable conditions this may be directly seen in the SAW propagation velocity and attenuation 
as function of magnetic field. This idea was spectacularly confirmed for the 2 DEG in GaAs heterojunctions \cite {Wixforth86,Willett90}. 
Extended investigations have shown that this method also allows uniquely to determine the wave-vector dependence of longitudinal 
conductivity \cite{Willett93}.

The 2DEG in  GaAs heterostructure  contains electrons with non-relativistic parabolic dispersion and hence a finite band mass. 
Recently graphene, a  single sheet honeycomb carbon layer,  has been identified as a genuine 2DEG system\cite{novoselov1}. It 
contains quasi-relativistic 
massless Dirac electrons with a linear dispersion around the Dirac point $K$ where the slope is given by the Fermi velocity 
$v_F=10^6$~m/s, obeying to
\begin{equation}
H=v_{F}(\sigma_xp_x+\sigma_yp_y),
\label{diracham}
\end{equation}
where the pseudospin variables ($\sigma$) arise from the two-sublattice structure. 
The 
transport and thermodynamic properties of this system has been much discussed and investigated, in particular 
in an external field \cite{geim,castro07}. They are distinctly different from the non-relativistic case which is mostly due to two 
effects: 
i) 
because of the linear dispersion the density of states (DOS) at the Dirac point $E=0$ vanishes according to $N(E)=|E|/D^2$ (per 
valley 
and spin; $D^2=2\pi v_F^2/A_c$, where $A_c$ is the area of the hexagonal cell). As a consequence the screening properties of Dirac 
electrons will be different from those of the nonrelativistic case \cite{wunsch,Hwang07}. ii) The Landau quantization in an external 
magnetic field 
leads to relativistic Landau levels at $E_\alpha(n)=\alpha\omega_c\sqrt{n+1}$  with $\alpha =\pm 1$, $n=0,1,2 ..$ and the 
Landau scale
 $\omega_c=v_F\sqrt{2e|B|}$. In addition there is a field independent level always at $E^*=0$.
%
%

There has been considerable discussion whether the Dirac point and the linear DOS in single layer graphene is stable against 
perturbations like buckling, disorder or excitonic gap formation. We ignore these subtle issues here and refer to the literature 
\cite{castro07}. Furthermore we note that the chemical potential $\mu$ may be easily moved away from the Dirac point by applying a 
gate 
voltage  $V_g$ via $V_g\sim n\sim \mu^2$\cite{brey} (with $n$ the carrier density). This is a new aspect as compared to the 
degenerate nonrelativistic 2DEG and we 
will employ it in our discussion of 
SAW propagation effects in graphene. Preliminary SAW experiments in zero field have been performed in Ref.~\onlinecite{Arsat09}. In 
this setup the aim was rather to detect surface adsorbates via their change of graphene electronic properties and ensuing SAW 
propagation anomalies. In the present work we want to look at intrinsic properties of graphene and investigate the field dependence, 
gate voltage (chemical potential) and wave vector dependence of the SAW propagation velocity and attenuation.
A possible experimental 
setup which we have in mind is shown schematically in Fig.~\ref{fig:Fig1}. The graphene sheet is sandwiched between two insulating 
piezoelectric layers with the same background dielectric constant $\epsilon$, similarly to Ref. \onlinecite{friedemann}. The SAW r.f. 
transducer and the detector are mounted on 
the top layer. Its thickness $d$ is assumed to be much smaller than the penetration depth of the SAW along $-z$. There may also be 
alternatives where the coupling of SAW in the piezoelectric substrate to the Dirac electrons happens purely capacitive across an 
empty space.

The analysis of SAW propagation in this setup has two distinct steps. Firstly the mechanical and electrodynamic boundary value 
problem for the SAW has to be solved which gives the SAW velocity change in terms of the (magneto-) conductive properties of the 
graphene sheet. This problem has been analyzed and solved in Ref.~\onlinecite{Simon96}. In a second step we calculate the 
magnetoconductivity of Dirac electrons as function of field and chemical potential which leads to the SAW velocity change 
and attenuation as function of field and gate voltage. In this part we follow the procedure described in 
Ref.~\onlinecite{doragraph}. 
Alternatively we consider the zero field case and investigate the wave vector (or frequency) dependence of SAW properties, which is 
related to the wave vector dependent ($B=0$) polarization given in Ref.~\onlinecite{Hwang07,wunsch}.

If one neglects the influence of the 2D relativistic electron gas the SAW propagation is a purely mechanical problem which is 
determined by  the elastic constants. We don't make any specific assumptions about the bulk and top layer material but rather 
treat it as an isotropic piezoelectric medium characterized by  longitudinal (L) and transverse (T) elastic constants or 
velocities $c_\beta$ or $v_\beta=\sqrt{(c_\beta/\rho)}$ ($\beta$ = L,T) respectively where $\rho$ is the mass density. 
The stress-free boundary condition for the geometry in Fig.~\ref{fig:Fig1} is $\sigma_{iz}=0$ ($i = x,y,z$). The solution 
of the boundary problem \cite{Landaubook} leads to the velocity of the surface Rayleigh waves $v_s$=$\xi v_T$. Here $v_s$  
is smaller than the transverse bulk velocity by a factor $0.87<\xi <0.95$ depending on the ratio $c_L/c_T$.
%
\begin{figure}
\includegraphics[width=7cm,height=11cm]{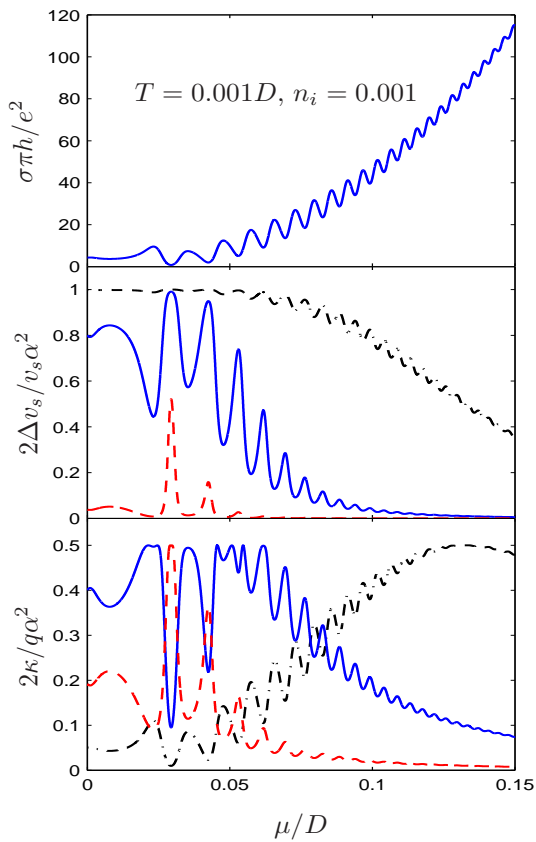}
\caption{Landau oscillations for constant field (corresponding to $N=1000$ LL within the cutoff frequency $D$) as function of 
the chemical potential $\mu$ tuned by the gate voltage for $T=0.001D$, $n_i=0.001$. Upper 
panel: longitudinal conductivity. Middle/lower panel: SAW velocity change/attenuation for $\sigma_m/\sigma_{xx}(\mu=0)=10$ 
(black dashed-dotted line), 
1 (blue solid line) and 1/10 (red dashed line).}
\label{fig:Fig2}
\end{figure}
%
Due to the piezoelectric coupling an external potential is created which couples to the 2DEG via the density response function.  
The latter is completely determined by conservation laws and linear response relations. The induced density in the 2DEG leads to 
an additional energy density which has to be added to the elastic energy density  causing a renormalization of the elastic constants 
or velocity. Because of the dissipative component in the density response the SAW velocity change is accompanied by an attenuation. 
It may be seen as Landau damping due to particle-hole excitations in the graphene sheet. By the above procedure \cite{Simon96} SAW 
velocity change $\Delta v_s$  and attenuation coefficient  $\kappa$ are obtained as 
\begin{equation}
\frac{\Delta v_s(q,B,\mu)}{v_s}-\frac{i\kappa(q,B,\mu)}{q}= \frac{\alpha(qd)^2/2}{1+i\sigma_{xx}(q,\omega,B,\mu)/\sigma_m}
\end{equation}       
Here $\omega =v_sq$ is the SAW frequency and ${\bf q}\parallel x$ the wave number. The penetration depths $\kappa^{-1}_L, 
\kappa^{-1}_T$ of displacement 
components $u_x, u_z$ into the bulk along $-z$ direction are of the order of the wave length $\lambda =2\pi/q$. The effective 
coupling 
$\alpha(qd)$ of SAW to the graphene 2DEG is a complicated expression composed of phenomenological parameters \cite{Simon96} which will 
not be given here. It depends nonmonotonically on $qd$ and in the limit $qd\ll 1$ or $\kappa_{(L,T)}q \ll 1$ assumed here is simply a 
constant to be determined by experiment. 

The central quantity is the longitudinal electronic conductivity $\sigma_{xx}(q,\omega;B,\mu)$ where $\omega = v_sq$. The normalization 
constant is $\sigma_m = v_s\epsilon_{eff}/2\pi$  with $\epsilon_{eff}=(\epsilon_0+\epsilon)/2$ for  $qd\ll 1$. The typical SAW frequency 
of $\sim$  0.1 GHz is extremely small ($\simeq 5\times 10^{-3}$~K) compared to the typical electron hole excitation energy (given by 
the chemical potential $\mu$  so 
that one may 
use the static limit $\sigma_{xx}(q;B,\mu)=\sigma_{xx}(q,\omega = 0;B,\mu)$). We will consider two different cases: i) finite 
field 
case in the presence of scatterers, which allows to neglect the possible $q$ dependence of the conductivity\cite{willettsurf}, if the 
scattering rate exceeds $v_Fq$ and ii) zero field ultraclean case. For the former, due to Landau quantization, oscillations in 
$\Delta v_s(B,\mu)$ will appear both as function of 
field and as function of the chemical potential $\mu(V_g)$ or gate voltage $V_g$. The latter is an attractive new feature peculiar 
for SAW in graphene but not for the non-relativistic 2DEG. One may keep the field constant and instead observe the Landau 
oscillations as function of gate voltage.
 %
\begin{figure}


\includegraphics[width=75mm]{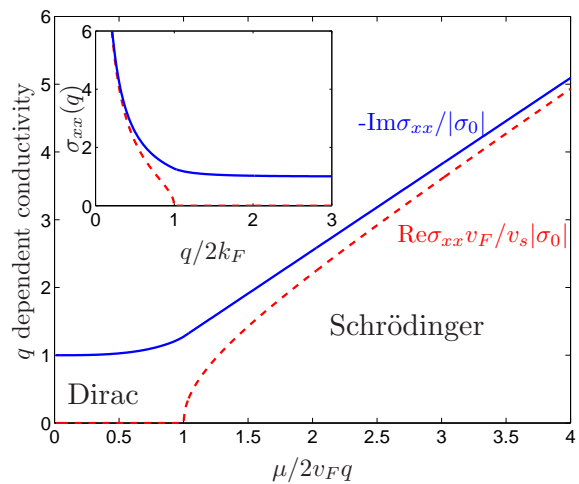}
\caption{Real (red dashed line) and imaginary (blue solid line) part of the conductivity for $B=0$ as function of chemical potential 
and wave number (inset) with $|\sigma_0|=e^2v_s/v_F$. The Dirac and Schr\"odinger like regions are separated at $2v_Fq=\mu$.}
\label{fig:Fig3}
\end{figure}
%
For the first case we need the field dependent conductivity $\sigma_{xx}(B,\mu)$ which was derived for localized disorder 
(vacancies representing the unitary limit\cite{peresalap}) in the
self-consistent Born approximation (SCBA) in Ref.~\onlinecite{doragraph}. The result is shown in the top panel of Fig.~\ref{fig:Fig2} 
as 
function of the chemical potential for a constant magnetic field for $\omega_c=D/\sqrt{N+1}$ with $N=1000$. Clearly the Landau 
oscillations in 
the conductivity also influence the SAW properties. 
When the chemical potential crosses a LL, the conductivity peaks, and $\Delta v_s$ develops a dip.
In between LL's, the conductivity is suppressed, causing the enhancement of the SAW velocity.
For large chemical potential when the graphene sheet becomes more 
metallic, both $\Delta v_s$ and $\kappa$ are suppressed.

The SAW velocity measurement in zero field presents a unique feature which cannot be realized by the usual transport 
measurements, which usually probe the conductivity at $\omega\gg v_Fq$, i.e. in the dynamic limit. Due to the finite SAW 
frequency 
the density response of the Dirac electrons may be probed at finite wave 
vector $q=\omega/v_s$ with $\omega\ll v_Fq$, thus probing the static properties of the electric response.
 Measurement of $\Delta v_s$ for various frequencies $\omega$ can therefore 
determine the density 
response of Dirac electrons for finite wave vector $\omega/v_s$. This has actually been proposed and carried out for the 
non-relativistic 2DEG in Ref.~\onlinecite{Willett93}.

The $q$ ($\parallel x$) and $\omega$ dependent longitudinal conductivity is calculated similarly to the polarization 
function\cite{wunsch,Hwang07},
and is given in the limit of $\omega=v_sq\ll (\mu,v_Fq)$ as
\begin{gather}
\sigma_{xx}(q)=\frac{4 v_s e^2}{\pi 
v_F}\left[\frac{v_s}{v_F}f_1\left(\frac{v_Fq}{2\mu}\right)+if_2\left(\frac{v_Fq}{2\mu}\right)\right],
\label{qcond}
\end{gather}
where
\begin{gather}
f_1(x)=\frac{\sqrt{1-x^2}}{x}\Theta(1-x),\\
f_2(x)=-\frac 1x+\frac {1} {2x}\left[\sqrt{1-\frac{1}{x^2}}-x\arccos\left(\frac 1x\right)\right]\Theta(x-1),
\end{gather}
and is shown in Fig. \ref{fig:Fig3}. It depends only on $v_Fq/\mu$, thus exhibits scaling behaviour.
For small frequencies, $\sigma_{xx}$ vanishes with $\omega$ as follows from Eq. \eqref{cont} via $\omega=v_sq$,
which explains the $v_s$ term on the r.h.s of Eq. \eqref{qcond}.
The real part of the conductivity, which determines the attenuation, is suppressed by the additional $v_s/v_F$ factor, 
thus the propagation velocity is more strongly influenced by the presence of graphene than the attenuation.

The density response or polarizability $\Pi(q)$ is related to the longitudinal conductivity by
\begin{equation}
\Pi(q_x,\omega)=\frac{iq_x^2}{\omega}\sigma_{xx}(q_x,\omega),
\label{cont}
\end{equation}
which follows from the charge continuity equation, and involves only the longitudinal component of the conductivity.
For ${\bf q}=(q_x,0)$, $q_x=q$.
Using the equation for the SAW velocity change and neglecting the attenuation we may easily obtain
\begin{equation}
\Pi(q)\simeq  q\sigma_m \left[\frac{\alpha^2}{2\Delta v_s(q)}-\frac{1}{v_s}\right]
\end{equation}
from the measured velocity change. The density response for monolayer graphene has also been calculated
in Ref.~\onlinecite{Hwang07,wunsch}, and SAW measurements can yield to the static polarizability as well.

\begin{figure}

\includegraphics[width=4cm]{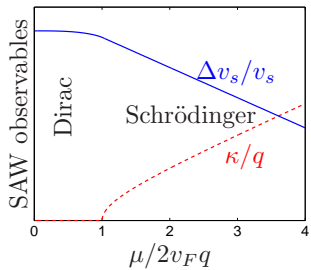}
\caption{Schematic view of the SAW velocity change (blue solid line) and attenuation (red dashed line) as a function of the 
chemical potential for $B=0$ clean graphene. For small $\mu$, the attenuation is negligible in the Dirac regime.}
\label{fig:Fig4}
\end{figure}

There are two limits to be considered
depending on whether $q$ is larger or smaller than twice the radius of the Fermi circle $k_F=\mu/v_F$.
For $2k_F>q$ what we refer to as Schr\"odinger like behaviour, $\sigma_{xx}=8v_se^2 k_F(\frac{v_s}{v_F}-i)/\pi v_F^2 q$, which 
holds 
true in a normal metal as well, and reveals the important $k_F/q$ dependence of the conductivity.
On the other hand, the peculiarity of graphene is the tunability of its Fermi energy with a gate voltage, which provides 
access to the $2k_F<q$ region, termed Dirac like regime, in which case 
$\sigma_0=-ie^2v_s/v_F$ exactly at the neutrality point ($\mu=0$).
Thus, there is practically no attenuation in this regime, only velocity renormalization.
Typical SAW wavelengths range from 2-150~$\mu$m\cite{paalanen}, which are translated to $v_Fq\sim 2-200$~K using $v_F$ for graphene. 
Graphene samples being ballistic over the micron-submicron scale imply a scattering rate around 50~K, thus, the 
above Dirac to Schr\"odinger like crossover can in principle be observed in clean samples with a tiny residual scattering rate at the 
neutrality point.
The resulting behaviour of SAW observables are shown in Fig. \ref{fig:Fig4}, which resembles closely to that of the bare 
conductivity. 
We also speculate that SAW at $\mu=0$ can identify the role inhomogeneities extending over several lattice constant (e.g. ripples, 
puddles) in graphene. When 
the SAW wavelength becomes comparable to their spatial extension, all waves passing through are scattered, 
and the sample is not transparent any more but acoustically opalescent.

In summary we have studied SAW propagation in monolayer graphene. 
By exploiting the chemical potential dependence of its conductivity, Landau oscillations are predicted in both the SAW 
velocity change and 
attenuation by varying the gate voltage, a feature which is only accessible due to the quasi-relativistic excitation spectrum in 
graphene, clearly absent in a 
normal 2DEG. SAW measurements in graphene can provide us with a deeper insight into the unconventional nature of the quantum Hall 
effect.
Without magnetic field, the SAW frequency changes measure directly the wavevector dependence of the longitudinal conductivity, which 
can reveal a Dirac to Schr\"odinger like crossover, with no attenuation in the Dirac regime. In addition, SAW propagation in pristine 
graphene can yield significant 
information about the dominant scattering mechanisms close to the neutrality point through $\sigma_{xx}(q_x)$.

\begin{acknowledgments}

We are grateful to M. Polini, A. Hill and J. Ebbecke for useful correspondences. 
This work was supported by the Hungarian
Scientific Research Funds under grant number K72613, and by the Bolyai program of the Hungarian Academy of Sciences
\end{acknowledgments}

\bibliographystyle{apsrev}
\bibliography{refgraph}

\end{document}